\documentstyle[12pt]{article}

\begin{document}

\author{M.\ Apostol  \\ 
Department of Theoretical Physics,\\Institute of Atomic Physics,
Magurele-Bucharest MG-6,\\POBox MG-35, Romania\\e-mail: apoma@theor1.ifa.ro}
\title{On metallic clusters squeezed in atomic cages }
\date{J.\ Theor.\ Phys.\ {\bf 9}\ (1995)}
\maketitle

\begin{abstract}
The stability of metallic clusters of sodium ($Na$) in the octahedral cages
of $Na$-doped fullerites $Na_6C_{60}$ and $Na_{11}C_{60}$ is discussed
within a Thomas-Fermi model. It is shown that the tetrahedral $Na_4$-cluster
in $Na_6C_{60}$ has an electric charge $\sim +2.7$ (in electron charge
units), while the body-centered cubic $Na_9$-cluster in $Na_{11}C_{60}$ is
almost electrically neutral.
\end{abstract}

The nature of the metallic clusters is currently receiving a great deal of
interest.\cite{Heer} Metallic clusters trapped in zeolites cages provide the
opportunity of studying them by various magnetic resonance techniques,\cite
{Harrison}$^{-}$\cite{Bifone} and, in this respect, the nature of the
chemical bonding, the distinction between metallic and covalent bondings and
the degree of ionicity of the clusters are of utmost importance. Recently,
the existence of $Na$-clusters in the octahedral cages of $Na$-doped $C_{60}$%
-fullerite has been suggested from $X$-ray analysis.\cite{Rosseinsky}$^{,}$%
\cite{Yildirim} Similar clusters of calcium ($Ca$) in $Ca$-doped $C_{60}$
has previously been pointed out.\cite{Kortan} Tetrahedral-shaped $Na_4$%
-clusters seem to build up in the octahedral cages of the (face-centered
cubic) $fcc$-$Na_6C_{60}$,\cite{Rosseinsky} with the $Na$ atoms disposed
alternately on the corners of a cube of side $2\;\AA $.\ Similarly,
body-centered cubes of $9$ sodium atoms ($Na_9$-cluster) seem to occur in
the octahedral cages of the $fcc$-$Na_{11}C_{60}$,\cite{Yildirim} with the
cube side $3.2\;\AA $. This is in contrast with the $C_{60}$-fullerite doped
with heavier species, such as $A_6C_{60}$, $A=K,\;Rb,\;Cs$, where the alkali
cations push through the $fcc$-structure of the host pristine $C_{60}$ and
distort it into a $bcc$-structure, while acquiring a conventional ionic
character.\cite{Zhou} The stability and the chemical bonding of isolated,
small-size alkali-clusters, like $Na$- and $Li$-clusters, have extensively
been studied.\cite{Boustani}$^{-}$\cite{Prince} Recently, the formation of $%
Na$-clusters in $Na_6C_{60}$ has been challenged, on computational grounds.%
\cite{Andreoni} Here we present a Thomas-Fermi model of $Na$-clusters in $%
Na_6C_{60}$ and $Na_{11}C_{60}$, discuss its stability and compute the
distribution of electric charges and the degree of ionization of the
clusters.

The main observation we start from is that the $Na$-$Na$ distance in these $%
Na$-clusters is much shorter than the $Na$-$Na$ distance $3.7\;\AA $ in bulk
metallic sodium, though slightly larger than twice the $Na^{+}$-radius $%
1.15\;\AA $. Indeed, the $Na$-$Na$ distance in the tetrahedral $Na_4$%
-cluster in $Na_6C_{60}$\cite{Rosseinsky} is $2\sqrt{2}\;\AA =2.82\;\AA $,
while in the cubic $Na_9$-cluster in $Na_{11}C_{60}$\cite{Yildirim} this
distance (the shortest) is $3.2\sqrt{3}/2\;\AA =2.75\;\AA $.\ This indicates
that, in comparison with the bulk metallic sodium, the $Na$-clusters in
doped $C_{60}$ are squeezed (or compressed) in the octahedral cages provided
by the neighbouring fullerene molecules. In the octahedral coordination of
the $fcc$-$C_{60}$ the fullerene molecules are placed at $\sim 7.1\;\AA $
from the coordination centre,\cite{Stephens}$^{,}$\cite{Murphy} and,
allowing for a radius of $\sim 3.5\;\AA $ of the fullerene molecule, one
obtains an average radius $R_0=2\;\AA $ of the octahedral cage.\cite
{Rosseinsky}$^{,}$\cite{Yildirim} For the tetrahedral $Na_4$-cluster, whose
atoms are placed at $R=2\sqrt{3}/2\;\AA =1.73\;\AA $ from the coordination
centre, this cage provides an average atomic radius of $2^{\frac 13}\;\AA $,
and an average inter-electron separation much smaller than the Bohr radius.\
This suggests that a Thomas-Fermi model is suitable for describing the $Na_4$%
-cluster. The confinement of the cluster in the octahedral cage is ensured
by the electronic clouds of (carbon) $C$-atoms on the surface of the
fullerene molecules, which, in order to be penetrated, require
higher-energetical electrons. They act as very high potential barriers. The
confinement of electrons in the much narrower tetrahedral sites of the $fcc$-%
$C_{60}$ (whose radius is estimated at $\sim 1.12\;\AA $\cite{Kortan}) may
certainly provide them with enough energy as to make them able of
penetrating the confinement, which may explain the absence of clusters in
these cages. Similar considerations apply for heavier-alkali atoms in the
octahedral cages, and even for the $Na_9$-cluster in $Na_{11}C_{60}$, whose
atoms, placed at $R=3.2\sqrt{3}/2\;\AA =2.75\;\AA $, clearly trespassed the $%
2\;\AA $ limit found above.\ However, the penetration of the electronic
clouds of the $C$-atoms proceeds gradually, and, in the octahedral cages,
the $C$-atoms provide another limit of about $4\;\AA $ in the $(111)$%
-directions: it is the distance from the coordination centre to the
octahedral faces passing through the centres of the fullerene molecules.\ We
may take, tentatively, a mean value of these two limits ($2\;\AA $ and $%
4\;\AA $), and set up an average cage radius $R_0\sim 3\;\AA $ for the $Na_9$%
-cluster. This gives an average atomic radius $3^{\frac 13}\;\AA $, which
again justifies a Thomas-Fermi model for this $Na_9$-cluster. It is
worth-remarking at this point that the resistance the cage walls oppose to
the penetrating electrons (which may also apply in the case of the zeolites
cages) is an expression of the Pauli exclusion principle, which requires
fastly-varying (in space) wavefunctions for the latter (and therefore higher
energy) if they are going to penetrate into the space regions already
occupied by atomic-core electrons. This confining effect of the cage
boundaries on an atomic cluster may be formalized as an external field
''pseudo-potential'', much in the same manner as it has already been done
for bulk metals.\cite{Chelikowsky} It can easily be seen that these
''pseudo-potentials'' would vary strongly over short distances (as in the
short-range parts of the inter-atomic Lennard-Jones or Born-Mayer
potentials, for example), and their functional dependence on distance (which
would depend, among others, on the number of confined electrons) will
provide an effective cage radius $R_0$ for a given cluster, as we shall
discuss below in more detail.

As it is well-known the Thomas-Fermi model is a qusi-classical theory of
interacting electrons and atomic nuclei which assumes that the Fermi
wavector $k_F$ of the electrons, and the electron density $n$,

\begin{equation}
\label{one}n=\frac 1{3\pi ^2}k_F^3\;\;, 
\end{equation}
vary slowly in space. This is true (except for distances very close to the
nuclei) due to the effective screening of the nuclear charges accomplished
by the high-density electrons. We shall see that this assumption is a
consistent one also in the present Thomas-Fermi model of clusters. Under
this assumption the particular shape of the atomic cluster is no longer very
relevant, so that we may introduce an additional simplification by assuming
that the total charge $z$ of the nuclei is uniformly distributed over a
spherical shell of radius $R$, plus a central $Na$-nucleus of charge $+11$
(in electron charge units) in the case of the $Na_9$-cluster. We denote by $%
V $ the electrostatic potential created by this distribution of positive
charges, and remark that its particular expression is not very important in
subsequent calculations; the only requirement on the positive charge
distribution is that of symmetry, from obvious reasons of stability, and the
approximation made is that of radial symmetry. In the high-density limit the
electron exchange interaction is a higher-order contribution (as well as
other contributions to the electron interaction energy), and we may assume
that an electrostatic potential $\varphi $ is generated by the electron
distribution, which does not depend on $k_F$.\ We may therefore set up the
basic Thomas-Fermi relationship (in atomic units Bohr radius $a_H=\hbar
^2/me^2=0.53\;\AA ,\;e^2/a_H=27.2\;eV$ and $\hbar =1$)

\begin{equation}
\label{two}\frac 12k_F^2-\varphi -V-U=-\varphi _0\;\;, 
\end{equation}
where $U$ is the potential well of the cage and $\varphi _0$ is the chemical
potential; the latter should be constant in order to ensure the local
equilibrium and equal to the ''pseudo-potential'' of the external field at
the cage frontier $R_0$, such as to prevent any flux of electrons pouring in
and out of the cage.\ The variation of this external ''pseudo-potential''
over the distance $R_0$ may be taken as the depth $U$ of the potential well.
Using $(1)$ and $(2)$ we may rewrite the Poisson equation $\Delta \varphi
=4\pi n$ as

\begin{equation}
\label{three}\Delta (\varphi +V+U-\varphi _0)=\frac{8\sqrt{2}}{3\pi }%
(\varphi +V+U-\varphi _0)^{\frac 32}\;, 
\end{equation}
everywhere except for $r=R$; introducing the reduced variables $x=r/R_0$ and

\begin{equation}
\label{four}\varphi +V+U-\varphi _0=\left( \frac{3\pi }{8\sqrt{2}R_0^2}%
\right) ^2\frac \chi x 
\end{equation}
we arrive at the Thomas-Fermi equation

\begin{equation}
\label{five}x^{\frac 12}\chi ^{^{\prime \prime }}=\chi ^{\frac 32}\;\;, 
\end{equation}
for any $x$ between $0$ and $1$ except for $x=a=R/R_0$. Since $n$ is a
continuous function $\varphi $ and its two first derivatives must be
continuous functions; it follows that the function $\chi $ has a slope
discontinuity (but is itself continuous) exactly as the derivative of $V$
has, {\it i.e}. at $x=a$; the magnitude of this slope jump will be obtained
below by Gauss$^{^{\prime }}$ law.\ It is therefore convenient to define $%
\chi _1(x)=\chi (x)$ for $0<x<a$ and $\chi _2(x)=\chi (x)$ for $a<x<1$, with
the continuity condition

\begin{equation}
\label{six}\chi _1\left( a\right) =\chi _2\left( a\right) \;\;. 
\end{equation}

The number of electrons inside a sphere of radius $x$ is easily obatined
from $(1)-(5)$ as

\begin{equation}
\label{seven}
\begin{array}{c}
N(x)=4\pi R_0^3\int_0^xdx\cdot x^2n= 
\frac{9\pi ^2}{128R_0^3}\int_0^xdx\left[ x\chi ^{^{\prime }}-\chi \right] =
\\ =\left\{ 
\begin{array}{c}
\frac{9\pi ^2}{128R_0^3}\left[ x\chi _1^{^{\prime }}-\chi _1+\chi
_1(0)\right] \;\;,\;\;0<x<a\;, \\ \frac{9\pi ^2}{128R_0^3}\left\{ x\chi
_2^{^{\prime }}-\chi _2+a\left[ \chi _1^{^{\prime }}(a)-\chi _2^{^{\prime
}}(a)\right] +\chi _1(0)\right\} \;\;,\;\;a<x<1\;, 
\end{array}
\right\} 
\end{array}
\end{equation}
and one remarks that it is continuous at $x=a$. The total number of
electrons in the cage is obtained from $(7)$ as

\begin{equation}
\label{eight}N=\frac{9\pi ^2}{128R_0^3}\left\{ \chi _2^{^{\prime }}(1)-\chi
_2(1)+a\left[ \chi _1^{^{\prime }}(a)-\chi _2^{^{\prime }}(a)\right] +\chi
_1(0)\right\} \;. 
\end{equation}
By Gauss$^{^{\prime }}$ law we get easily the net charge inside the sphere
of radius $x$ as

\begin{equation}
\label{nine}q(x)=\frac{9\pi ^2}{128R_0^3}(\chi -x\chi ^{^{\prime }})\;\;, 
\end{equation}
whence, by using $(7)$, we obtain the positive charge

\begin{equation}
\label{ten}q_{1+}(x)=\frac{9\pi ^2}{128R_0^3}\chi _1(0) 
\end{equation}
inside the sphere of radius $x<a$. We remark that, if it exists, it is
concentrated at $x=0$. Obviously, for the $Na_4$-cluster we should require 
\begin{equation}
\label{eleven}\chi _1(0)=0\;\;, 
\end{equation}
while for the body-centered cubic $Na_9$-cluster the corresponding boundary
condition is

\begin{equation}
\label{twelve}\frac{9\pi ^2}{128R_0^3}\chi _1(0)=11\;. 
\end{equation}
The total charge inside the sphere of radius $x$, $a<x<1$, given by $(9)$
consists of $z$-positive charges distributed over the spherical shell of the
cluster, a charge $+11$ at the centre, if the case, and $-N(x)$ electron
charges; whence, by using $(7)$, we get the magnitude 
\begin{equation}
\label{thirteen}z=\frac{9\pi ^2}{128R_0^3}a\left[ \chi _1^{^{\prime
}}(a)-\chi _2^{^{\prime }}(a)\right] 
\end{equation}
in the slope discontinuity of the function $\chi $, as expected; $(13)$
provides the third boundary condition for our problem, with $z=44$ for the $%
Na_4$-cluster and $z=88$ for the $Na_9$-cluster. The fourth boundary
condition needed for the uniqueness of the solution of $(5)$ is provided by
the stability condition of the positive shell of the cluster. It is easy to
get the electric self-field acting on the shell; in the absence of the
central atom it is given by 
\begin{equation}
\label{fourteen}E_s=z/2R^2\;\;, 
\end{equation}
while in the case of the centered cluster we obtain

\begin{equation}
\label{fifteen}E_s=z/2R^2+11/R^2\;\;. 
\end{equation}
The electric field on the shell due to the electrons is readily obtained as

\begin{equation}
\label{sixteen}E_e=-\frac{\partial \varphi }{\partial r}\mid _{r=R}=\frac{%
9\pi ^2}{128R_0^5}\frac 1{a^2}\left[ \chi _1(a)-a\chi _1^{^{\prime
}}(a)-\chi _1(0)\right] 
\end{equation}
(or equivalently in terms of $\chi _2$), and we remark that, in contrast to $%
(14)$ and $(15)$, it is negative, {\it i.e.} the electrons try to stabilize
the exploding cluster shell. Using $(13)$ we arrive easily to the
equilibrium condition

\begin{equation}
\label{seventeen}2\chi _1(a)=a\left[ \chi _1^{^{\prime }}(a)+\chi
_2^{^{\prime }}(a)\right] 
\end{equation}
both for the non-centered and for the centered cluster.\ The equation $(5)$
is solved numerically with the boundary conditions given by $(6),\;(11)-(13)$
and $(17),$ both for the $Na_4$- and the $Na_9$-cluster, and the results are
described below. In each case we have been interested especially in the
total charge

\begin{equation}
\label{eighteen}q=\frac{9\pi ^2}{128R_0^3}\left[ \chi _2(1)-\chi
_2^{^{\prime }}(1)\right] 
\end{equation}
in the cage, as given by $(9)$.

In the case of the $Na_4$-cluster we have used $R=1.73\;\AA $ and $%
R_0=2\;\AA \;(a=R/R_0=0.86)$, as suggested by the $X$-ray diffraction data.%
\cite{Rosseinsky}$^{,}$\cite{Kortan} The function $\chi $ is obtained by
solving numerically the above equations; it corresponds to a total charge $%
q=+2.72$, {\it i.e.} an average charge $+0.68$ per each $Na$-ion. As we have
discussed above, the numerical results obtained for the function $\chi $ are
valid everywhere except for a range of about twice the Bohr radius around
the position of the positive shell of charges, where the variation of $\chi
/x$ is too large; even so, one can estimate that almost half of the
electrons are located there, ensuring thus the screening of the positive
charges. Another limitation of the present calculations occurs from the
assumption of radial symmetry for the function $\chi $. We may estimate the
error introduced by this approximation as follows. Assuming the same
variation of the electron density per unit length in all directions the
relative contribution of the angular part of the variation of the potential
derivative with respect to the radial one is $\delta \varphi _a^{^{\prime
}}/\delta ^{^{\prime }}\varphi _r\sim \sqrt{\frac{4\pi }s}$, for a cluster
consisting of $s$ atoms; on the other hand, as these small variations are
proportional to the variations of the distance we have $\left( \delta
\varphi _r^{^{\prime }}\right) ^2+2\left( \delta \varphi _a^{^{\prime
}}\right) ^2=\left( \delta \varphi _r^{^{\prime }}\right) _0^2$, where the
latter corresponds to the neglect of the angular part. As the variations of
the potential derivatives are also proportional to the charge (from the
Poisson equation), we find then easily that the relative error made in
estimating the charge is at most $\sim 1-\sqrt{1+8\pi /s}/(1+2\sqrt{4\pi /s})
$.\ In the case of the $Na_4$-cluster this amounts to$\sim 40\%$; one can
see that the radial-symmetry approximation improves its accuracy for large $s
$.

Increasing the amount of negative charge on the cluster the position of the
maximum value of $\chi $ moves toward smaller values of $x$ (smaller values
of $R$), and it exhibits an opposite behaviour on increasing the degree of
ionization. The variation of $\chi $ with respect to the cage radius $R_0$
brings into discussion the energy of the cluster. One can check easily that
the density of the electronic kinetic energy is given by

\begin{equation}
\label{nineteen}\varepsilon _{kin}=\frac 1{2\pi ^2}\int_0^{k_F}dk\cdot k^4=%
\frac{9\left( 3\pi /32\right) ^3}{10R_0^{10}}\cdot \frac{\chi ^{\frac 52}}{%
x^{\frac 52}}\;\;^{,} 
\end{equation}
where $(2)$ and $(4)$ have been used; whence one can obtain
straightforwardly the total kinetic energy of the electrons

\begin{equation}
\label{twenty}{\cal E}_{kin}=\frac{3\left( 3\pi /8\right) ^4}{20R_0^7}%
\int_0^1dx\cdot \frac{\chi ^{\frac 52}}{x^{\frac 12}}\;\;^{.} 
\end{equation}
We obtain similarly the potential energy of the electrons

\begin{equation}
\label{twentyone}{\cal E}_{pot}=-\frac{\left( 3\pi /8\right) ^4}{4R_0^7}%
\int_0^1dx\cdot \frac{\chi ^{\frac 52}}{x^{\frac 12}}-N\varphi _0\;\;^{,} 
\end{equation}
so that the electronic energy may be expressed as ${\cal E}_{el}={\cal E}%
_1-N\varphi _0$, where 
\begin{equation}
\label{twentytwo}{\cal E}_1=-\frac{\left( 3\pi /8\right) ^4}{10R_0^7}%
\int_0^1dx\cdot \frac{\chi ^{\frac 52}}{x^{\frac 12}}\;\;. 
\end{equation}
The self-energy of the positive charges is given by ${\cal E}_s={\cal E}%
_0-zU $ for the non-centered cluster and ${\cal E}_s={\cal E}_0-\left(
z+11\right) U$ for the centered one, where ${\cal E}_0=z^2/2R$ in the former
case and ${\cal E}_0=z^2/2R+11z/R$ in the latter. On the other hand,
expressing the interaction energy of the shell with the electrons in two
distinct ways, namely, $z\varphi \left( R\right) =-\int d{\bf r\;}nV$, we
get 
\begin{equation}
\label{twentythree}U-\varphi _0=\frac{9\pi ^2}{128R_0^4}\chi _2^{^{\prime
}}\left( 1\right) \;\;. 
\end{equation}
Summing up all these contributions we obtain the total energy of the cluster
as

\begin{equation}
\label{twentyfour}{\cal E}={\cal E}_0+{\cal E}_1+{\cal E}_2-\left(
2z-q\right) \varphi _0\;\;\;, 
\end{equation}
where ${\cal E}_2=-z\left( U-\varphi _0\right) $ as given above; in the case
of the centered cluster we should replace $z$ by $z+11$ in $\left( 24\right) 
$. The numerical integration gives ${\cal E}_1=$ $-28$ and ${\cal E}_2=-42$
for the $Na_4$-cluster, while ${\cal E}_0=296$; the total energy, consisting
of $(24)$ and the energy of the electrons transferred to the cage walls,
must be negative, in order the cluster be formed. For small values of the
total charge of the cluster the latter contribution to the total energy is
small, so that we may estimate that the energy $E$ given by $(24)$ must
acquire negative values. Hence it follows that $\varphi _0$ must be of the
order of $(296-28-42)/N\sim 2.6$ (for $N=44$). On the other hand, $\varphi
_0 $ equals the external ''pseudo-potential'' at the cage frontier $R_0$, as
we have discussed before, whence one may have an estimation of the magnitude
of this ''pseudo-potential''. One can see, therefore, that it acquires high
values, as to ensure the confinement of the squeezed cluster within the $R_0$
range. In addition, as the present estimations are valid over a scale length
larger than the Bohr radius, one can also see that the external
''pseudo-potential'' has a high variation over short distances. This
illustrates again the gradual resistance developed by the inner-core
electrons of the environment to the penetrating ''pressure'' of the confined
cluster. The electron density varies also fastly over short distances at the
frontier of the cage, in order to ensure the delimitation of the compressed
cluster from its confining environment. The ''pressure'' exerted by the
cluster electrons on their environment results in the deformation of the
electronic clouds of the surrounding atoms. Having determined $\varphi _0$
by the external ''pseudo-potential'' at the cage frontier $R_0$, the two
remaining parameters $R_0$and $R$ will be obtained, within a complete
theory, by minimizing the total energy of the cluster (given by $(24)$) and
the electrons transferred to the cage walls.\ Of course, this requires first
the computation of the ''pseudo-potential''.

From the discussion above one can see that the results of the computation
are very sensitive to the values of $R$ and $R_0$; this sensitivity may
result in a higher inaccuracy for clusters consisting of a greater number of
atoms. This can be seen in the case of the $Na_9$-cluster, where, according
to the experimental data suggested by the $X$-ray analysis,\cite{Yildirim}
we have used $R=2.75\;\AA $ and $R_0=3.15\;\AA $ ($a=R/R_0=0.87$). The
function $\chi $ is obtained again by solving the corresponding equations
given above; it corresponds, within errors of $\pm 0.03\;\AA $ in $R$ and $%
R_0$, to a total charge of the cluster $q\sim \pm 1$.\ At this level of
approximation we may conclude that the centered-cubic $Na_9$-cluster is
practically neutral. The same discussion regarding the screening and the
value of the chemical potential $\varphi _0$ ($\sim 2.4$) applies here as
for the $Na_4$-cluster. It is worth-noting that a similar calculation for a
non-centered cubic cluster consisting of $8$ atoms of $Na$, gives a total
charge $q\sim +1$, for the same values of $R$ and $R_0.$ Finally, noticing
that the theory contains the charge $z$ only in the product $zR_0^3$ (see,
for example, $(12),\;(13)\;$and $(18)$) we can say that similar clusters
formed by lithium may also form in the octahedral cages of the fullerite
with a smaller size, or, equivalently, with a diminished cluster-size and
ionization charge.

In conclusion one may say that the calculations presented here indicate a
metallic bonding of the $Na_4$- and $Na_{11}$-clusters squeezed into the
octahedral cages of the $Na$-fullerides. We emphasize that this squeezing is
a central feature of the picture presented here, whose direct consequence is
that of affecting all the electron states in the cluster; enough electronic
energy may be gained in this way as to stabilize the cluster at lower values
of $R$, as compared with $R_0$. This is in contrast, at least for the $Na_4$%
-cluster, with recent claims\cite{Andreoni} that the $Na$-atoms in $%
Na_6C_{60}$ are fully ionized and that the octahedral $Na$-ions are placed
much further away from the coordination centre. These claims, based on
numerical simulations within the local-density approximation, make use of a
certain distinction between the valence and the core electrons, which might
be inappropriate for the squeezed clusters. As we have seen from the present
calculations, almost half of the electrons seem to be affected by the
presence of the neighbouring atoms in the cluster, which indicates a drastic
change in the inner electronic states of the isolated atoms. The metallic
character of the alkali clusters in fullerides may be tested by various
techniques, most notably by Raman and $^{23}Na$-NMR spectroscopies.\cite{com}
In addition, it is worth-mentioning that the metallic clusters squeezed in
atomic cages may exhibit two types of collective modes, associated, one,
with the vibrations of the positively-charged ions and, another, with the
oscillations of the electron density. Preliminary calculations indicate that
these two modes are in the range of the ultra-violet optical spectroscopy
and the electron spectroscopy.

\end{document}